\documentclass[aps,prd,onecolumn,notitlepage,nofootinbib,showpacs,showkeys,superscriptaddress,groupedaddress]{revtex4-2}
\usepackage[T1]{fontenc}
\usepackage[utf8]{inputenc}
\usepackage{bm}
\usepackage{mathtools}
\usepackage{esint}
\usepackage{xcolor}
\usepackage{array}
\usepackage{float}
\usepackage{graphicx}
\PassOptionsToPackage{normalem}{ulem}
\usepackage{ulem}
\usepackage{hyperref}
\usepackage{multirow}
\begin{document}
	
	\title{Covariant 3+1 correspondence of the spatially covariant gravity and the degeneracy conditions}

	\author{Yu-Min Hu}%
	\affiliation{%
	School of Physics and Astronomy, Sun Yat-sen University, Guangzhou 510275, China}

	\author{Xian Gao}%
	\email[Corresponding author: ]{gaoxian@mail.sysu.edu.cn}
	\affiliation{%
	School of Physics and Astronomy, Sun Yat-sen University, Guangzhou 510275, China}

	\date{November 17, 2021}
	
	\begin{abstract}
		A necessary condition for a generally covariant scalar-tensor theory to be ghostfree is that it contains no extra degrees of freedom in the unitary gauge, in which the Lagrangian corresponds to the spatially covariant gravity. Comparing with analysing the scalar-tensor theory directly, it is simpler to map the spatially covariant gravity to the generally covariant scalar-tensor theory using the gauge recovering procedures. In order to ensure the resulting scalar-tensor theory to be ghostfree absolutely, i.e., no matter if the unitary gauge is accessible, a further covariant degeneracy/constraint analysis is required. We develop a method of covariant 3+1 correspondence, which map the spatially covariant gravity to the scalar-tensor theory in 3+1 decomposed form without fixing any coordinates. Then the degeneracy conditions to remove the extra degrees of freedom can be found easily. As an illustration of this approach, we show how the Horndeski theory is recovered from the spatially covariant gravity. This approach can be used to find more general ghostfree scalar-tensor theory.
	\end{abstract}

\maketitle


\section{Introduction}

The scalar-tensor theory is widely studied as one of the alternatives of the general relativity (GR), which introduces additional scalar degree(s) of freedom (DOF) other than the two tensorial DOF’s (i.e., the gravitational waves) of the GR.
In the theoretical aspects, one of the central problems in the developments of scalar-tensor theory is to introduce only the healthy DOF's while evading the ghostlike (or simply the unwanted) DOF's that are associated with the Ostrogradsky instabilities \cite{Woodard:2015zca,Motohashi:2020psc}.

The most straightforward approach is to construct a generally covariant Lagrangian, in which the scalar field(s) is (are) coupled to the spacetime metric covariantly.
This is actually what the name scalar-tensor theory is referred to originally.
In the past decade, the successful construction of the higher derivative single field scalar-tensor theory with a single scalar DOF has significantly enlarged our scope of the scalar-tensor theory \cite{Horndeski:1974wa,Nicolis:2008in,Deffayet:2011gz,Kobayashi:2011nu,Gleyzes:2014dya,Gleyzes:2014qga,Langlois:2015cwa,BenAchour:2016fzp,Crisostomi:2016czh}.
Ghostfree generally covariant scalar-tensor theory with higher derivatives can be constructed by finely tuning  the higher derivatives such that the higher derivatives are degenerate (see \cite{Langlois:2018dxi,Kobayashi:2019hrl} for reviews and \cite{Motohashi:2014opa,Klein:2016aiq,Motohashi:2016ftl,Motohashi:2017eya,Motohashi:2018pxg} for general discussions the degeneracy conditions).
Nevertheless, the generally covariant approach becomes more and more involved when going to higher orders both in the derivatives of the scalar field and in the curvature.

From the point of view of DOF's, the scalar-tensor theory can be understood as any effective gravitational theory that propagates the tensor as well as the scalar DOF's.
In particular, a class of pure metric theories that respect only the spatial diffeomorphism was proposed and shown to have two tensor DOF's with an additional scalar DOF \cite{Gao:2014soa,Gao:2014fra}.
In this sense, the ghost condensation \cite{ArkaniHamed:2003uy}, the effective theory of inflation \cite{Cheung:2007st,Gubitosi:2012hu} as well as the Ho\v{r}ava gravity \cite{Horava:2009uw,Horava:2008ih} can be viewed as sub-classes of the spatially covariant gravity, although which were proposed originally by different motivations.
In particular, the degeneracy can be made easily, even trivially, in the spatially covariant gravity description, not only because the Lagrangian is built directly in a spacetime split manner, but also because the Lagrangian gets simplified dramatically when fixing the unitary gauge. 
In fact, one may try even ambitiously to build theories respecting only the spatial covariance at the level of the Hamiltonian instead of the Lagrangian \cite{Aoki:2018zcv,Aoki:2018brq,Mukohyama:2019unx,Yao:2020tur}.

These two apparently different approaches to the scalar-tensor theory are related by the ``gauge fixing/recovering'' procedures.
If the gradient of the scalar field is timelike, we may fix the time coordinate as the scalar field $t=\phi$ such that the resulting theory appears to be a theory of spatially covariant gravity.
Conversely, starting from a spatially covariant gravity, we may derive the corresponding generally covariant Lagrangian of the scalar field and spacetime metric by the so-called Stueckelberg trick\footnote{This is also to perform a broken time diffeomorphism.}.
A natural idea is thus we first build the ghostfree spatially covariant graivty, and then map it to the generally covariant scalar-tensor theory, which yields the scalar-tensor theory that appears to be ghostfree at least in the unitary gauge.
Based on this idea, both the generally covariant and spatially covariant monomials have been classified and their correspondence has been investigated in \cite{Gao:2020juc,Gao:2020yzr,Gao:2020qxy}.

There are at least two subtleties in this correspondence.
Firstly, the reversibility of this gauge fixing/recovering procedures relies on the assumption of a timelike scalar field.
Secondly, even we assume that the scalar field is timelike, the generally covariant scalar-tensor theory got from the spatially covariant gravity appears arguably to have extra unwanted DOF's in coordinates that are not adapted to the unitary gauge \cite{DeFelice:2018ewo,DeFelice:2021hps}\footnote{Such an extra mode is dubbed ``instantaneous'' or ``shadowy'' mode since it propagates with an infinite speed. See also \cite{Gabadadze:2004iv,Blas:2010hb,Blas:2011ni} for early discussions.}.
In order to construct the scalar-tensor theory that is ghostfree ``absolutely'', i.e., no matter whether the scalar field is timelike or not and in any coordinates, one needs to perform a further degeneracy or constraint analysis. 
Usually this is done by making a 3+1 decomposition and performing the constraint analysis in the Hamiltonian formalism.

Comparing with finding the degeneracy conditions for the most general scalar-tensor theory directly (e.g., the approach taken in \cite{Langlois:2015cwa,BenAchour:2016fzp,Crisostomi:2016czh}), starting from the spatially covariant gravity has already saved works a lot.
However, one still needs two steps, by first finding the generally covariant scalar-tensor theory that corresponds to the ghostfree spatially covariant gravity, and then making a degeneracy analysis which needs a further covariant 3+1 decomposition.
One may wonder that if we can derive the covariant 3+1 correspondence of the spatially covariant gravity directly.
This work is devoted to this issue.

Generally, there are three apparently different formulations of the scalar-tensor theory. One is the generally covariant scalar-tensor theory, of which the Lagrangian is built of the scalar field coupled to the metric through generally covariant derivatives.
The second is the spatially covariant gravity, which corresponds to the generally covariant scalar-tensor theory in the coordinates adapted to the unitary gauge.
The last one is the generally covariant 3+1 decomposition of the scalar-tensor theory, which is convenient to be used for the covariant degeneracy/constraint analysis.
In this work, we shall develop a formalism, which we dub the ``covariant 3+1 correspondence'', that can be used to derive the explicit generally covariant 3+1 expressions from the spatially covariant gravity.

This work is organized as follows.
In Sec. \ref{sec:3faces} we describe the three formulations of the scalar-tensor theory and their correspondences.
In Sec. \ref{sec:covcorr} we derive the explicit expressions of the covariant 3+1 correspondence.
We apply this correspondence in Sec. \ref{sec:d2}, in which we derive the covariant 3+1 correspondence of the spatially covariant gravity of $d=2$ with $d$ the total number of derivatives in spatially covariant gravity formulation.
By cancelling all the dangerous terms, we determine the degeneracy conditions easily. 
In Sec. \ref{sec:d3} and Sec. \ref{sec:d3a}, we further apply this method to spatially covariant gravity of $d=3$ without and with the acceleration, respectively.
No surprisingly, we can recover the whole Lagrangian of the Horndeski theory easily by this method.
We summarize our results in Sec. \ref{sec:con}.

\section{Three faces of the scalar-tensor theory} \label{sec:3faces}

\subsection{Generally covariant formulations}

The generally covariant scalar-tensor theory (GST) is usually referred to the theory of scalar field(s) coupled to the spacetime metric.
In the present work, we concentrate on the case of a single scale field.
The action takes the general form
\begin{equation}
	S_{\mathrm{GST}}=\int\mathrm{d}^{4}x\sqrt{-g}\,\mathcal{L}\left(\phi;g_{ab},\varepsilon_{abcd},\,{}^{4}\!R_{abcd};\nabla_{a}\right), \label{S_GST}
\end{equation}
in which the Lagrangian is built of the scalar field $\phi$, the spacetime metric $g_{ab}$, the spacetime curvature tensor ${}^{4}\!R_{abcd}$ as well as their covariant derivatives.
The possible parity violation is encoded in the 4-dimension Levi-Civita tensor $\varepsilon_{abcd}$. 
It is the scalar-tensor theory in the form of (\ref{S_GST}), in which the general covariance is manifest, that is the subject in \cite{Horndeski:1974wa,Nicolis:2008in,Deffayet:2011gz,Kobayashi:2011nu,Gleyzes:2014dya,Gleyzes:2014qga,Langlois:2015cwa,BenAchour:2016fzp,Crisostomi:2016czh} and also used in practical model buildings of cosmology and black holes, etc..

For the purpose of degeneracy/constraint analysis, splitting the 4 dimensional objects into their temporal and spatial parts, i.e., the so-called 3+1 decomposition, is needed.
The starting point of the 3+1 decomposition is a timelike vector field $n_{a}$ with normalization $n_{a}n^{a}=-1$.
As usual, this timelike vector field is assumed to be hypersurface orthogonal, and thus the induced metric which projects any tensor field on the spatial hypersurface is
\begin{equation}
	h_{ab} \equiv g_{ab}+n_{a}n_{b}.
\end{equation}
All the 4 dimensional quantities are then split into parts that are orthogonal and tangent to the spatial hypersurface by projecting with $n^{a}$ and $h_{ab}$, respectively. 
The decomposition of the 4 dimensional curvature tensor yields the Gauss-Codazzi-Ricci equations.
For the scalar field,  we have
\begin{equation}
	\nabla_{a}\phi=-n_{a}\pounds_{\bm{n}}\phi+\mathrm{D}_{a}\phi,\label{nabla_phi_dec}
\end{equation}
where $\pounds_{\bm{n}}$ stands for the Lie derivative with respect to $n^{a}$, $\mathrm{D}_{a}$ is the projected derivative defined by
\begin{equation}
	\mathrm{D}_{a}\phi := h_{a}^{\phantom{a}a'}\nabla_{a'}\phi,
\end{equation}
which is also the covariant derivative compatible with $h_{ab}$.
The decompositions of the second and the third order derivatives of the scalar field with respect to a general normal vector $n^{a}$ can be found in \cite{Gao:2020yzr}.

With these settings, we can derive the covariant 3+1 decomposition (COD) of any 4 dimensional quantities. The GST action (\ref{S_GST}) can be recast in the form
	\begin{equation}
		S_{\mathrm{COD}}=\int\mathrm{d}^{4}x\, \sqrt{-g}\, \mathcal{L}\left(\phi;n_{a},h_{ab},\varepsilon_{abcd},{}^{3}\!R_{ab};\mathrm{D}_{a},\pounds_{\bm{n}}\right). \label{S_COD}
	\end{equation}
We emphasize that the action (\ref{S_COD}) is generally covariant since $n_{a}$ is an arbitrary hypersurface orthogonal unit timelike vector field, and we have not yet chosen any specific coordinates.
In particular, the familiar lapse function $N$ and shift vector $N^{a}$ do not appear in the Lagrangian\footnote{They merely encode the gauge freedom of choosing the time and space directions, i.e., fixing the coordinates.}.
In (\ref{S_COD}), ${}^{3}\!R_{ab}$ is the intrinsic curvature of the hypersurfaces.
The projected derivative $\mathrm{D}_{a}$ and the Lie derivative $\pounds_{\bm{n}}$ can be viewed as the ``intrinsic'' and ``extrinsic'' derivatives, respectively.
The Lie derivatives of $n^{a}$ and $h_{ab}$ 
	\begin{eqnarray}
		a_{a} & = & \pounds_{\bm{n}}n_{a}, \label{a_def_n}\\
		K_{ab} & = & \frac{1}{2}\pounds_{\bm{n}}h_{ab}, \label{K_def_n}
	\end{eqnarray} 
define the acceleration and the extrinsic curvature as usual.

\subsection{Spatially covariant formulation}

In the action (\ref{S_COD}), $n_{a}$ is an arbitrary unit timelike vector field that is hypersurface orthogonal. 
While the scalar field $\phi$ itself specifies a foliation of hypersurfaces with $\phi = \mathrm{const.}$.
In particular, when the gradient of the scalar field is also timelike, we are allowed to choose $n_{a} = u_{a}$, where
	\begin{equation}
		u_{a} \equiv -\frac{1}{\sqrt{2X}}\nabla_{a}\phi, \label{ua_def}
	\end{equation}
with the canonical kinetic term of the scalar field $X = -\frac{1}{2}\nabla_{a}\phi\nabla^{a}\phi$.
$u_{a}$ is nothing but the normal vector of the hypersurfaces with constant $\phi$, which satisfies the normalization $u_{a}u^{a}=-1$.
Choosing $n_{a} = u_{a}$ corresponds to the so-called unitary gauge in the literature\footnote{Usually the ``unitary gauge'' is referred to fixing the time coordinate $t=\phi$ in the literature. In this work, for the purpose of distinguishing the generally covariant and spatially covariant formulations, we use ``unitary gauge'' to denote choosing $n_{a} = u_{a}$. In particular, no specific coordinates have been fixed.}.

In the unitary gauge, i.e., when being decomposed with respect to the foliation specified by the scalar field $\phi$ itself, the decompositions of the derivatives of the scalar field get dramatically simplified. 
All the spatial derivatives of the scalar field drop out since
\begin{equation}
	\overset{\mathrm{u}}{\mathrm{D}}_{a}\phi\equiv\overset{\mathrm{u}}{h}_{a}{}^{a'}\nabla_{a'}\phi=0,
\end{equation}
where $\overset{\mathrm{u}}{h}_{ab}$ is defined by
\begin{equation}
	\overset{\mathrm{u}}{h}_{ab} \equiv  g_{ab}+u_{a}u_{b} .\label{Hab_ug_sf}
\end{equation}
Here and throughout this paper, an overscript ``$\mathrm{u}$'' denotes quantities defined with respect to $u_{a}$ \cite{Gao:2020yzr}, which is related to the scalar field through (\ref{ua_def}).
The first order derivative of the scalar field (\ref{nabla_phi_dec}) is thus written as $\nabla_a \phi = -u_{a}/N$, where we introduce
\begin{equation}
	\frac{1}{N}=\sqrt{2X}=\pounds_{\bm{u}}\phi.\label{lapse_ug_sf}
\end{equation}
In (\ref{lapse_ug_sf}) $N$ is nothing but the lapse function, which arises since we have identified the ``space'' to be the hypersurfaces of constant $\phi$.
The decompositions of the second and the third order derivatives of the scalar field in the unitary gauge can be found in \cite{Gao:2014soa,Gao:2014fra,Gao:2020yzr}.
Replacing $n_{a}$ by $u_{a}$ in the action (\ref{S_COD}) yields
	\begin{equation}
		S_{\mathrm{u.g.}}=\int \mathrm{d}^{4}x\, \sqrt{-g}\, \mathcal{L}\left(\phi,u_{a},\overset{\mathrm{u}}{h}_{ab},\varepsilon_{abcd},{}^{3}\!\overset{\mathrm{u}}{R}_{ab};\overset{\mathrm{u}}{\mathrm{D}}_{a},\pounds_{\bm{u}}\right). \label{S_ugauge}
	\end{equation}
At this point, all the ingredients are generally covariant. As a result, the unitary gauge action (\ref{S_ugauge})  is generally covariant.

In the unitary gauge, since $n_{a}$ is chosen to be $u_{a}$, the coordinates that are adapted to the foliation, i.e., the Arnowitt-Deser-Misner (ADM) coordinates, correspond to fixing  $t= \phi$ (while spatial coordinates are left free). 
In these particular coordinates, we have $u_{a} = -N \delta_{a}^{0}$ and the time direction $t^{a} = \delta^{a}_{0}$.
The unitary gauge action (\ref{S_ugauge}) is recast to 
	\begin{equation}
	S_{\mathrm{SCG}}=\int \mathrm{d}t\mathrm{d}^{3}x\, N\sqrt{h}\, \mathcal{L}\left(t,N,h_{ij},\varepsilon_{ijk},{}^{3}\!R_{ij};\nabla_{i},\pounds_{\bm{u}}\right), \label{S_SCG}
\end{equation}
where $\pounds_{\bm{u}}$ is now understood to be $\frac{1}{N}\left(\partial_{t}-\pounds_{\vec{N}}\right)$ with $\vec{N}$ the spatial component of $t^{a}-Nu^{a} = \left(0,N^{i}\right)$.
Since the time coordinate $t$ is fixed to be the value of $\phi$, the general covariance is broken to the spatial diffeomorphism.
The action (\ref{S_SCG}) appears to be a pure metric theory respecting spatial covariance, which we dub the spatially covariant gravity (SCG).
The effective theory of inflation \cite{Cheung:2007st,Gubitosi:2012hu}, the Ho\v{r}ava gravity \cite{Horava:2009uw,Horava:2008ih} as well as the more general framework proposed in \cite{Gao:2014soa,Gao:2014fra} can be viewed as sub-classes of the general action of SCG (\ref{S_SCG}),

\subsection{Theory triangle: relations among different formulations} \label{sec:scg2st}

We have now three apparently different formulations of the theory.
From the point of view of keeping the general covariance manifestly and/or of making the spacetime decomposition explicitly, different formulations have their own merits. 
\begin{itemize}
	\item The generally covariant scalar-tensor theory (GST) (\ref{S_GST}) :\\
	The general covariance is manifest in the action of GST, which is also convenient for model buildings in the cosmology and black hole physics. However, more calculations are needed to derive its spacetime decomposition in order to make the degeneracy/constraint analysis.
	\item The spatially covariant gravity (SCG)  (\ref{S_SCG}):\\
	The SCG is written in the already spacetime-decomposed manner, which is convenient for controlling the number of DOF's through a strict degeneracy/constraint analysis. In particular, comparing with the GST, the degenerate SCG Lagrangian with the desired number of DOF's can be constructed much easier. For example, the SCG \cite{Gao:2014soa,Gao:2014fra} contains only the extrinsic curvature as the kinetic terms and thus is trivially degenerate. SCG with the a dynamical lapse function has also been investigated in \cite{Gao:2018znj,Gao:2019lpz,Lin:2020nro} (see also \cite{Motohashi:2020wxj}). However, the general covariance is explicitly broken in SCG.
	\item The covariant 3+1 decomposition  (COD) (\ref{S_COD}):\\
	The COD Lagrangian can be viewed as the balance between GST and SCG. It is written in the spacetime-decomposed form and thus is convenient to perform the constraint analysis. On the other hand, it is generally covariant and has the exact equivalence to the GST. In other words, the Lagrangians of COD and GST are exactly the the same, but merely written in different forms.
\end{itemize}

\begin{figure}[H]
	\begin{centering}
		\includegraphics[scale=0.8]{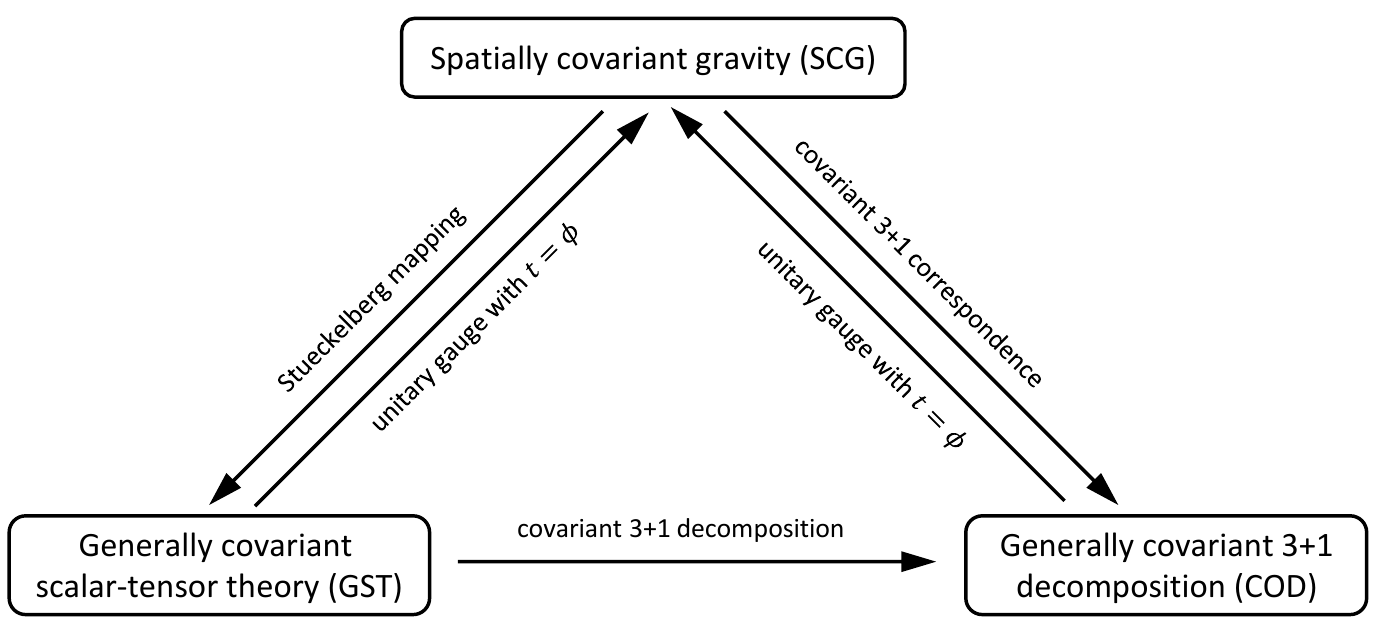}
		\par\end{centering}
	\caption{Theory triangle: three faces of the scalar-tensor theory.}
	\label{fig:covdec}
\end{figure}

The relations among the three formulations are depicted in Fig. \ref{fig:covdec}.
Starting from the GST, we get the COD by performing a covariant 3+1 decomposition. Then we arrive at the SCG by choosing the unitary gauge and fixing the time coordinate.
With this approach, the Lagrangian of the Horndeski theory in the unitary gauge was derived in \cite{Gleyzes:2013ooa}.
Similar analysis was performed to get a geometric reformulation of the quadratic degenerate higher-order scalar-tensor theory \cite{Langlois:2020xbc}.
For our purpose to use the SCG to generate GST theories, the inverse procedures of the 3+1 decomposition and the gauge fixing are required.
To this end, we must determine the GST quantities that correspond to the SCG quantities.
This procedure has been used in the covariant formulation of the Ho\v{r}ava gravity \cite{Germani:2009yt,Blas:2009yd,Jacobson:2010mx,Blas:2010hb} (see also \cite{Chagoya:2018yna,Barausse:2021dza}), and is sometimes dubbed the Stueckelberg trick.

Since the SCG quantities are simply the unitary gauge quantities after fixing the time coordinate $t=\phi$, while the later are the GST quantities after choosing the unitary gauge $n_{a}= u_{a}$, the one-to-one correspondence between a SCG expression and a GST expression can be easily set up. 
For example, (\ref{ua_def}) and (\ref{lapse_ug_sf}) can be viewed as the GST correspondences of $u_{a}$ and $N$, respectively.
The extrinsic curvature corresponds to
\begin{equation}
	K_{ij} \rightarrow \overset{\mathrm{u}}{K}_{ab} =  -\frac{1}{\sqrt{2X}}\overset{\mathrm{u}}{h}_{aa'}\,\overset{\mathrm{u}}{h}_{bb'}\nabla^{a'}\nabla^{b'}\phi,\label{Kuab_sf}
\end{equation}
where $\overset{\mathrm{u}}{h}_{ab}$ is defined in (\ref{Hab_ug_sf}), which now should be understood as
\begin{equation}
	\overset{\mathrm{u}}{h}_{ab} = g_{ab}+\frac{1}{2X}\nabla_{a}\phi\nabla_{b}\phi.\label{huab_sf}
\end{equation}
By plugging (\ref{huab_sf}) in (\ref{Kuab_sf}), we get the GST correspondence of $K_{ij}$.
We refer to \cite{Gao:2020yzr} for the more complete and detailed correspondences between the GST and SCG expressions.

As we have argued before, since the degenerate SCG Lagrangian can be constructed much easier than GST, one may use the degenerate SCG as the ``seed theory'', and map it to the space of GST theories using the above correspondence.
The resulting theory is the GST theory that is ghostfree, or propagates the correct number of DOF's, when the unitary gauge is accessible\footnote{Scalar-tensor theory with this property is also referred to be ``U-degenerate'', i.e., being degenerate in the unitary gauge \cite{DeFelice:2018ewo}.}.
In fact, this has already been performed for the GST and SCG polynomials \cite{Gao:2020qxy} from the linear algebraic point of view.

When the unitary gauge is not accessible, or at least when we do not fix the time coordinate to be the scalar field, apparently there arise extra DOF's which might be ghostlike.
Our final purpose is to obtain the GST theory that is ghostfree ``absolutely'', which has the correct number of DOF's in the generally covariant sense no matter whether the scalar field is timelike so that the unitary gauge is accessible or not, and shows no extra DOF's in arbitrary coordinates.
To this end, a further covariant 3+1 decomposition is inevitable, which results in the COD formulation of the GST.
This ``two-step'' approach, i.e., SCG$\rightarrow$GST$\rightarrow$COD, although is correct and straightforward, is technically involved since both steps involves complicated correspondences among expressions in different formulations.

The main purpose of this work is to find a ``one-step'' approach, i.e., a method to derived the  COD expressions from the SCG expressions directly, which we dub the covariant 3+1 correspondence and shall explain in the next section.

\section{Covariant 3+1 correspondence} \label{sec:covcorr}

The covariant 3+1 correspondence is conceptually simple, which combines the above two steps  together, but without expanding the intermediate GST in terms of the scalar field and 4 dimensional geometric quantities explicitly.

Firstly, we covariantize the SCG expressions by determining the corresponding unitary gauge expressions.
For example, the spatial metric $h_{ij}$, although appears to be 3-dimension tensor, is actually the spatial component of a 4-dimension tensor 
\begin{equation}
	h_{ij} \rightarrow \overset{u}{h}_{ab} = g_{ab} + u_{a}u_{b},
\end{equation}
where $u_{a}$ is nothing but the normalized gradient of the scalar field (\ref{ua_def}).
Secondly, instead of recasting the unitary gauge expressions in terms of the scalar field and 4-dimension geometric quantities explicitly (e.g., (\ref{huab_sf})), we make a further 3+1 decomposition with respect to a general spacelike foliation with normal vector $n_{a}$.
For $u_{a}$,  we write
\begin{equation}
	u_{a}=-n_{a}\alpha+\beta_{a}, \label{ua_dec}
\end{equation}
and require that $n^{a}\beta_{a}\equiv 0$.
Since both $u_{a}$ and $n_{a}$ are normalized (with sign $-1$), $\alpha$ and $\beta_{a}$ are not independent, which satisfy
\begin{equation}
	\alpha=-\sqrt{1+\beta^{2}},
\end{equation}
where $\beta^2 \equiv \beta_{a}\beta^{a}$.
Since $u_{a}$ is given in (\ref{ua_def}), $\alpha$ and $\beta_{a}$ are related to the derivatives of the scalar field by
\begin{eqnarray}
	\alpha & =- & \frac{\pounds_{\bm{n}}\phi}{\sqrt{2X}},\label{sf_u_alpha}\\
	\beta_{a} & = & -\frac{\mathrm{D}_{a}\phi}{\sqrt{2X}},\label{sf_u_beta}
\end{eqnarray}
where the canonical kinetic term $X$ is now decomposed to be
\begin{equation}
	X=\frac{1}{2}\left(\pounds_{\bm{n}}\phi\right)^{2}-\frac{1}{2}\mathrm{D}_{a}\phi\mathrm{D}^{a}\phi.
\end{equation}
Throughout this paper, quantities without any overscript are defined with respect to a general normal vector field $n_{a}$.
Therefore (\ref{ua_dec}) becomes
\begin{equation}
	u_{a}=n_{a}\sqrt{1+\beta^{2}}+\beta_{a}. \label{ua_beta_rel_gen}
\end{equation}
(\ref{ua_beta_rel_gen}) is the starting point of the following analysis, which is nothing but the covariant 3+1 decomposition of the normalized gradient of the scalar field without fixing any coordinates.
One can see from (\ref{ua_beta_rel_gen}) that $\beta_{a}$ encodes the deviation of the general foliation from the foliation specified by the scalar field. Therefore the unitary gauge is simply defined to be
\begin{equation}
	\text{unitary gauge:}\quad  \beta_{a} \rightarrow 0 ,
\end{equation}
which implies $n_{a} \rightarrow u_{a}$ as expected.

The covariant 3+1 correspondence of the spatial metric is\footnote{Throughout this paper, symmetrization is normalized, e.g., $A_{(a}B_{b)} \equiv \frac{1}{2}(A_{a}B_{b}+A_{b}B_{a})$.}
\begin{equation}
	\overset{u}{h}_{ab}=n_{a}n_{b}\overset{u}{h}_{\bm{n}\bm{n}}-2n_{(a}\overset{u}{h}_{\hat{b})\bm{n}}+\overset{u}{h}_{\hat{a}\hat{b}}, \label{indH_u_dec_abb}
\end{equation}
where
\begin{eqnarray}
	\overset{u}{h}_{\bm{n}\bm{n}} & = & \beta^{2},\label{indh_u_nn}\\
	\overset{u}{h}_{\hat{a}\bm{n}} & = & \alpha\beta_{a},\label{indh_u_hn}\\
	\overset{u}{h}_{\hat{a}\hat{b}} & = & h_{ab}+\beta_{a}\beta_{b}.\label{indh_u_hh}
\end{eqnarray}
Here $h_{ab}$ is the induced metric associated with $n_{a}$, i.e., $h_{ab}\equiv g_{ab} + n_{a}n_{b}$.
Here and in what follows, we use the notation in \cite{Deruelle:2012xv} that for a general spacetime tensor, an index replaced by $\bm{n}$ denotes contraction with $n_{a}$,  and indices with a hat denote projection with $h_{ab}$, i.e.,
	\begin{equation}
		T_{\cdots\bm{n}\cdots}=n^{a}T_{\cdots a\cdots}, \quad T_{\cdots\hat{a}\cdots}=h_{a}^{\phantom{a}a'}T_{\cdots a'\cdots}.
	\end{equation}
From (\ref{indH_u_dec_abb}) it is clear that the difference of $\overset{u}{h}_{ab}$ and $h_{ab}$ is completely encoded in the non-vanishing $\beta_{a}$.
Therefore $\overset{u}{h}_{ab} \rightarrow h_{ab}$ in the unitary gauge.

In the following, we derive the explicit expressions in the covariant 3+1 correspondence.
The fundamental objects are the covariant derivatives of $u_{a}$. For the first order derivative of $u_{a}$, we have
	\begin{equation}
		\nabla_{a}u_{b}=n_{a}n_{b}A-n_{a}B_{b}-\tilde{B}_{a}n_{b}+\Delta_{ab},\label{nabla_u_dec}
	\end{equation}
with
	\begin{eqnarray}
		A\equiv\nabla_{\bm{n}}u_{\bm{n}} & = & \dot{\alpha}-a^{c}\beta_{c},\label{nabla_u_dec_nn}\\
		B_{b}\equiv\nabla_{\bm{n}}u_{\hat{b}} & = & -a_{b}\alpha+\dot{\beta}_{b}-K_{b}^{c}\beta_{c},\label{nabla_u_dec_nh}\\
		\tilde{B}_{a}\equiv\nabla_{\hat{a}}u_{\bm{n}} & = & \mathrm{D}_{a}\alpha-K_{a}^{c}\beta_{c},\label{nabla_u_dec_hn}\\
		\Delta_{ab}\equiv\nabla_{\hat{a}}u_{\hat{b}} & = & -K_{ab}\alpha+\mathrm{D}_{a}\beta_{b}.\label{nabla_u_dec_hh}
	\end{eqnarray}
Throughout this work, overdots on the spatial tensors with lower indices denote Lie derivatives with respect to the general normal vector $n^{a}$, e.g., $\dot{\alpha} = \pounds_{\bm{n}}\alpha$, $\dot{\beta}_{a} \equiv \pounds_{\bm{n}} \beta_{a}$, $\ddot{\beta}_{a} \equiv \pounds_{\bm{n}}^{2} \beta_{a}$, etc..
Occasionally we also use dotted spatial tensors with upper indices for shorthand, in which the upper indices are raised by the inverse induced metric $h^{ab}$, e.g., $\dot{\beta}^{a} \equiv h^{ab}\dot{\beta}_{b}$, $\dot{K}^{ab}\equiv h^{aa'}h^{bb'}\dot{K}_{a'b'}$, etc.\footnote{Therefore $\dot{\beta}^{a} \equiv h^{aa'} \pounds_{\bm{n}}\beta_{a'}\neq \pounds_{\bm{n}}\beta^{a}$.}.
Evaluating the  Lie derivative of (\ref{sf_u_beta})  explicitly yields
\begin{equation}
	\dot{\beta}_{a}=-\frac{1}{2X}\beta_{a}\dot{X}-\frac{1}{\sqrt{2X}}\left(\mathrm{D}_{a}\dot{\phi}+a_{a}\dot{\phi}\right),
\end{equation}
where
\begin{equation}
	\dot{X}=\ddot{\phi}\dot{\phi}-\mathrm{D}^{a}\phi\left(\mathrm{D}_{a}\dot{\phi}+a_{a}\dot{\phi}\right)+K^{ab}\mathrm{D}_{a}\phi\mathrm{D}_{b}\phi.\label{LieD_n_X_xpl_fin}
\end{equation}
From (\ref{LieD_n_X_xpl_fin}) it is transparent that $\dot{\beta}_{a}$ contains the second order Lie derivative of the scalar field $\ddot{\phi}$ through $\dot{X}$, which should be degenerate (with the extrinsic curvature) in order not to excite the unwanted DOF's.

When considering the third order derivative of the scalar field, the second order derivative of $u_{a}$ will arise. We have
	\begin{eqnarray}
		\nabla_{c}\nabla_{a}u_{b} & = & -n_{c}n_{a}n_{b}U+n_{c}n_{a}V_{b}+n_{c}n_{b}\tilde{V}_{a}+n_{a}n_{b}W_{c}\nonumber \\
		&  & -n_{c}X_{ab}-n_{a}Y_{cb}-n_{b}\tilde{Y}_{ca}+Z_{cab},\label{nabla2_u_dec}
	\end{eqnarray}
	with
	\begin{eqnarray}
		U & = & \dot{A}-a^{d}B_{d}-a^{d}\tilde{B}_{d},\\
		V_{b} & = & -a_{b}A+\dot{B}_{b}-B_{d}K_{b}^{d}-\Delta_{db}a^{d},\\
		\tilde{V}_{a} & = & -a_{a}A+\dot{\tilde{B}}_{a}-\tilde{B}_{d}K_{a}^{d}-\Delta_{ad}a^{d},\\
		W_{c} & = & \mathrm{D}_{c}A-K_{c}^{d}B_{d}-K_{c}^{d}\tilde{B}_{d},\\
		X_{ab} & = & -a_{a}B_{b}-\tilde{B}_{a}a_{b}+\dot{\Delta}_{ab}-\Delta_{ad}K_{b}^{d}-\Delta_{db}K_{a}^{d},\\
		Y_{cb} & = & -K_{cb}A+\mathrm{D}_{c}B_{b}-K_{c}^{d}\Delta_{db},\\
		\tilde{Y}_{ca} & = & -K_{ca}A+\mathrm{D}_{c}\tilde{B}_{a}-K_{c}^{d}\Delta_{ad},\\
		Z_{cab} & = & -K_{ca}B_{b}-\tilde{B}_{a}K_{cb}+\mathrm{D}_{c}\Delta_{ab},
	\end{eqnarray}
where $A,B_{b},\tilde{B}_{a},\Delta_{ab}$ are given in (\ref{nabla_u_dec_nn})-(\ref{nabla_u_dec_hh}).
For later convenience, we also evaluate the Lie derivatives of $A,B_{b},\tilde{B}_{a},\Delta_{ab}$ explicitly, which are given by	
	\begin{equation}
		\dot{A}=\ddot{\alpha}-\beta^{b}\dot{a}_{b}-a^{b}\dot{\beta}_{b}+2K^{ab}a_{b}\beta_{a},
	\end{equation}
	\begin{equation}
		\dot{B}_{b}=-\alpha\dot{a}_{b}-a_{b}\dot{\alpha}+\ddot{\beta}_{b}-\beta^{c}\dot{K}_{bc}-K_{b}^{d}\dot{\beta}_{d}+2K_{bc}K^{cd}\beta_{d},
	\end{equation}
	\begin{equation}
		\dot{\tilde{B}}_{a}=\mathrm{D}_{a}\dot{\alpha}+a_{a}\dot{\alpha}-\beta^{c}\dot{K}_{ac}-K_{a}^{d}\dot{\beta}_{d}+2K_{ac}K^{cd}\beta_{d},
	\end{equation}
and
	\begin{eqnarray}
		\dot{\Delta}_{ab} & = & -\alpha\dot{K}_{ab}-K_{ab}\dot{\alpha}+\mathrm{D}_{a}\dot{\beta}_{b}+a_{a}\dot{\beta}_{b}\nonumber \\
		&  & -\left(a_{a}K_{bd}+a_{b}K_{da}-a_{d}K_{ab}\right)\beta^{d}\nonumber \\
		&  & -\left(\mathrm{D}_{a}K_{bd}+\mathrm{D}_{b}K_{da}-\mathrm{D}_{d}K_{ba}\right)\beta^{d}.
	\end{eqnarray}

We are ready to use (\ref{nabla_u_dec}) and (\ref{nabla2_u_dec}) to derive the covariant 3+1 correspondences of various geometric quantities.
For the extrinsic curvature, it is convenient to use the expression
\begin{equation}
	\overset{u}{K}_{ab}=\overset{u}{h}_{a}{}^{a'}\overset{u}{h}_{b}{}^{b'}\nabla_{(a'}u_{b')}.\label{Kab_u_def}
\end{equation}
It immediately follows that
\begin{equation}
	\overset{u}{K}_{ab}=n_{a}n_{b}\overset{u}{K}_{\bm{n}\bm{n}}-2n_{(a}\overset{u}{K}_{\hat{b})\bm{n}}+\overset{u}{K}_{\hat{a}\hat{b}},\label{Kab_u_dec}
\end{equation}
where
\begin{eqnarray}
	\overset{u}{K}_{\bm{n}\bm{n}} & = & -\beta^{2}\frac{1}{\alpha}\beta^{c}\dot{\beta}_{c}-\frac{1}{\alpha}K^{cd}\beta_{c}\beta_{d}+\beta^{2}a^{c}\beta_{c}+\beta^{c}\beta^{d}\mathrm{D}_{(c}\beta_{d)},\label{Kab_u_nn_xpl}
\end{eqnarray}
\begin{eqnarray}
	\overset{u}{K}_{\hat{a}\bm{n}} & = & \frac{1}{2}\beta_{a}\left(-\beta^{c}\dot{\beta}_{c}+\alpha\beta^{c}a_{c}+\frac{1}{\alpha}\beta^{c}\beta^{d}\mathrm{D}_{(c}\beta_{d)}\right)\nonumber \\
	&  & -\frac{1}{2}\beta^{2}\dot{\beta}_{a}+\frac{1}{2}\alpha\beta^{2}a_{a}-K_{ad}\beta^{d}+\frac{1}{2}\frac{1}{\alpha}\beta^{d}\mathrm{D}_{a}\beta_{d}+\frac{1}{2}\alpha\beta^{d}\mathrm{D}_{d}\beta_{a},\label{Kab_u_hn_xpl}
\end{eqnarray}
and
\begin{eqnarray}
	\overset{u}{K}_{\hat{a}\hat{b}} & = & -K_{ab}\alpha+\mathrm{D}_{(a}\beta_{b)}-\frac{1}{2}\beta_{a}\left(\alpha\dot{\beta}_{b}-a_{b}\alpha^{2}-\beta^{c}\mathrm{D}_{c}\beta_{b}\right)\nonumber \\
	&  & -\frac{1}{2}\beta_{b}\left(\alpha\dot{\beta}_{a}-a_{a}\alpha^{2}-\beta^{c}\mathrm{D}_{c}\beta_{a}\right).\label{Kab_u_hh_xpl}
\end{eqnarray}

For the acceleration, we shall use the expression
\begin{equation}
	\overset{u}{a}_{a}\equiv u^{b}\nabla_{b}u_{a}.
\end{equation}
It follows that
\begin{equation}
	\overset{u}{a}_{a}=-n_{a}\overset{u}{a}_{\bm{n}}+\overset{u}{a}_{\hat{a}},\label{acc_u_dec}
\end{equation}
where
\begin{equation}
	\overset{u}{a}_{\bm{n}}=-\beta^{c}\dot{\beta}_{c}+\alpha a^{c}\beta_{c}+\frac{1}{\alpha}\beta^{b}\beta^{c}\mathrm{D}_{b}\beta_{c},\label{acc_u_dec_n}
\end{equation}
and
\begin{equation}
	\overset{u}{a}_{\hat{a}}=-\alpha\dot{\beta}_{a}+a_{a}\alpha^{2}+\beta^{b}\mathrm{D}_{b}\beta_{a}.\label{acc_u_dec_h}
\end{equation}

For the spatial Ricci tensor, we make use of
	\begin{equation}
		\,{}^{3}\!\overset{u}{R}_{ab}=\overset{u}{h}_{a}{}^{a'}\overset{u}{h}_{b}{}^{b'}\overset{u}{h}{}^{cd}\overset{u}{\mathcal{R}}_{a'cb'd},\label{Ricci3d_u_Gauss_con}
	\end{equation}
where $\overset{u}{\mathcal{R}}_{acbd}$ is defined to be
	\begin{equation}
		\overset{u}{\mathcal{R}}_{acbd}=\,{}^{4}\!R_{acbd}-\nabla_{(a}u_{b)}\nabla_{(c}u_{d)}+\nabla_{(a}u_{d)}\nabla_{(c}u_{b)}.\label{calRu_DuDu}
	\end{equation}
Note $\overset{u}{\mathcal{R}}_{acbd}$ has exactly the same (anti-)symmetries of the spacetime Riemann tensor. Therefore there are 3 independent projections with $n_{a}$ and $h_{ab}$. By using the Gauss-Codazzi-Ricci equations of the Riemann tensor and (\ref{nabla_u_dec}), we find
	\begin{eqnarray}
		\overset{u}{\mathcal{R}}_{\hat{c}\bm{n}\hat{d}\bm{n}} & = & -\dot{K}_{cd}+K_{ce}K_{d}^{e}+a_{c}a_{d}+\mathrm{D}_{c}a_{d}\nonumber \\
		&  & -\left(-K_{cd}\alpha+\mathrm{D}_{(c}\beta_{d)}\right)\left(\dot{\alpha}-a^{e}\beta_{e}\right)\nonumber \\
		&  & +\frac{1}{4}\left(\dot{\beta}_{c}-a_{c}\alpha-2K_{c}^{e}\beta_{e}+\mathrm{D}_{c}\alpha\right)\left(\dot{\beta}_{d}-a_{d}\alpha-2K_{d}^{f}\beta_{f}+\mathrm{D}_{d}\alpha\right),\label{calRu_hnhn}
	\end{eqnarray}
	and
	\begin{eqnarray}
		\overset{u}{\mathcal{R}}_{\hat{a}'\hat{c}\hat{d}\bm{n}} & = & \mathrm{D}_{a'}K_{cd}-\mathrm{D}_{c}K_{a'd}\nonumber \\
		&  & -\frac{1}{2}\left(-K_{a'd}\alpha+\mathrm{D}_{(a'}\beta_{d)}\right)\left(\dot{\beta}_{c}-a_{c}\alpha-2K_{c}^{e}\beta_{e}+\mathrm{D}_{c}\alpha\right)\nonumber \\
		&  & +\frac{1}{2}\left(-K_{cd}\alpha+\mathrm{D}_{(c}\beta_{d)}\right)\left(\dot{\beta}_{a'}-a_{a'}\alpha-2K_{a'}^{e}\beta_{e}+\mathrm{D}_{a'}\alpha\right),\label{calRu_hhhn}
	\end{eqnarray}
	and
	\begin{eqnarray}
		\overset{u}{\mathcal{R}}_{\hat{a}'\hat{c}\hat{b}'\hat{d}} & = & \,{}^{3}\!R_{a'cb'd}+\left(K_{a'b'}K_{dc}-K_{a'd}K_{b'c}\right)\nonumber \\
		&  & -\left(-K_{a'b'}\alpha+\mathrm{D}_{(a'}\beta_{b')}\right)\left(-K_{cd}\alpha+\mathrm{D}_{(c}\beta_{d)}\right)\nonumber \\
		&  & +\left(-K_{a'd}\alpha+\mathrm{D}_{(a'}\beta_{d)}\right)\left(-K_{cb'}\alpha+\mathrm{D}_{(c}\beta_{b')}\right).\label{calRu_hhhh}
	\end{eqnarray}
Plugging (\ref{calRu_DuDu}) together with the above projections in (\ref{Ricci3d_u_Gauss_con}), after long and tedious manipulations, we find
	\begin{equation}
		^{3}\!\overset{u}{R}_{ab}=n_{a}n_{b}\,{}^{3}\!\overset{u}{R}_{\bm{n}\bm{n}}-2n_{(a}\,{}^{3}\!\overset{u}{R}_{\hat{b})\bm{n}}+\,{}^{3}\!\overset{u}{R}_{\hat{a}\hat{b}},\label{Rab_u_dec}
	\end{equation}
	where
	\begin{eqnarray}
		^{3}\!\overset{u}{R}_{\bm{n}\bm{n}} & = & \beta^{2}\left(\beta^{2}h^{cd}-\beta^{c}\beta^{d}\right)\overset{u}{\mathcal{R}}_{\hat{c}\bm{n}\hat{d}\bm{n}}\nonumber \\
		&  & +2\beta^{2}\alpha\beta^{a'}h^{cd}\overset{u}{\mathcal{R}}_{\hat{a}'\hat{c}\hat{d}\bm{n}}\nonumber \\
		&  & +\alpha^{2}\beta^{a'}\beta^{b'}h^{cd}\overset{u}{\mathcal{R}}_{\hat{a}'\hat{c}\hat{b}'\hat{d}},
	\end{eqnarray}
	and
	\begin{eqnarray}
		^{3}\!\overset{u}{R}_{\hat{b}\bm{n}} & = & \beta_{b}\alpha\left(\beta^{2}h^{cd}-\beta^{c}\beta^{d}\right)\overset{u}{\mathcal{R}}_{\hat{c}\bm{n}\hat{d}\bm{n}}\nonumber \\
		&  & +\left[h_{b}^{\phantom{b}a'}\left(\beta^{2}h^{cd}-\beta^{c}\beta^{d}\right)+\left(1+2\beta^{2}\right)\beta_{b}\beta^{a'}h^{cd}\right]\overset{u}{\mathcal{R}}_{\hat{a}'\hat{c}\hat{d}\bm{n}}\nonumber \\
		&  & +\left(h_{b}^{\phantom{b}b'}+\beta_{b}\beta^{b'}\right)\alpha\beta^{a'}h^{cd}\overset{u}{\mathcal{R}}_{\hat{a}'\hat{c}\hat{b}'\hat{d}},
	\end{eqnarray}
	and
	\begin{eqnarray}
		^{3}\!\overset{u}{R}_{\hat{a}\hat{b}} & = & \left[\beta_{a}\beta_{b}\left(\alpha^{2}h^{cd}-\beta^{c}\beta^{d}\right)+\beta^{2}h_{a}^{\phantom{a}c}h_{b}^{\phantom{a}d}-h_{a}^{\phantom{a}c}\beta_{b}\beta^{d}-h_{b}^{\phantom{a}d}\beta_{a}\beta^{c}\right]\overset{u}{\mathcal{R}}_{\hat{c}\bm{n}\hat{d}\bm{n}}\nonumber \\
		&  & -\alpha\left[h_{a}^{\phantom{b}a'}h_{b}^{\phantom{b}d}\beta^{c}+h_{b}^{\phantom{b}a'}h_{a}^{\phantom{b}d}\beta^{c}-\left(h_{a}^{\phantom{b}a'}\beta_{b}+h_{b}^{\phantom{b}a'}\beta_{a}\right)h^{cd}-2\beta_{a}\beta_{b}\beta^{a'}h^{cd}\right]\overset{u}{\mathcal{R}}_{\hat{a}'\hat{c}\hat{d}\bm{n}}\nonumber \\
		&  & +\left[h_{a}^{\phantom{a}a'}h_{b}^{\phantom{a}b'}\left(h^{cd}+\beta^{c}\beta^{d}\right)+\left(h_{a}^{\phantom{a}a'}\beta_{b}+h_{b}^{\phantom{a}a'}\beta_{a}\right)\beta^{b'}h^{cd}+\beta_{a}\beta_{b}\beta^{a'}\beta^{b'}h^{cd}\right]\overset{u}{\mathcal{R}}_{\hat{a}'\hat{c}\hat{b}'\hat{d}},
	\end{eqnarray}
where $\overset{u}{\mathcal{R}}_{\hat{c}\bm{n}\hat{d}\bm{n}}$, $\overset{u}{\mathcal{R}}_{\hat{a}'\hat{c}\hat{d}\bm{n}}$
and $\overset{u}{\mathcal{R}}_{\hat{a}'\hat{c}\hat{b}'\hat{d}}$ are given in (\ref{calRu_hnhn})-(\ref{calRu_hhhh}), respectively.

For the purpose to analyse the scalar-tensor theory involving the third order derivative of the scalar field, we also need the covariant 3+1 correspondence of the spatial derivatives of the extrinsic curvature and of the acceleration.
It is convenient to employ the expression
	\begin{equation}
		\overset{u}{\mathrm{D}}_{c}\overset{u}{K}_{ab}=\overset{u}{h}_{c}{}^{c'}\overset{u}{h}_{a}{}^{a'}\overset{u}{h}_{b}{}^{b'}\overset{u}{\mathcal{K}}_{c'a'b'},
	\end{equation}
with
	\begin{equation}
		\overset{u}{\mathcal{K}}_{cab}=\nabla_{c}\nabla_{(a}u_{b)}+\nabla_{c}u_{(a|}u^{d}\nabla_{d}u_{|b)}.
	\end{equation}
Together with (\ref{nabla_u_dec}) and (\ref{nabla2_u_dec}), we can get the covariant 3+1 correspondence of $\overset{u}{\mathrm{D}}_{c}\overset{u}{K}_{ab}$ explicitly. Similarly, we make use of
	\begin{equation}
		\overset{u}{\mathrm{D}}_{a}\overset{u}{a}_{b}=\overset{u}{h}_{a}{}^{a'}\overset{u}{h}_{b}{}^{b'}\overset{u}{\mathcal{A}}_{a'b'},
	\end{equation}
with
	\begin{equation}
		\overset{u}{\mathcal{A}}_{ab}=u^{c}\nabla_{a}\nabla_{c}u_{b}+\nabla_{a}u^{c}\nabla_{c}u_{b}.
	\end{equation}
Together with (\ref{nabla_u_dec}) and (\ref{nabla2_u_dec}), we then get the covariant 3+1 correspondence of $\overset{u}{\mathrm{D}}_{a}\overset{u}{a}_{b}$ explicitly.

Before proceeding, let us take the trace of the extrinsic curvature $K$ as an illustrative example. From (\ref{Kab_u_dec}) one finds
	\begin{eqnarray}
		K\rightarrow\overset{u}{K} & \equiv & g^{ab}\overset{u}{K}_{ab}\nonumber \\
		& = & \sqrt{1+\beta^{2}}K-\frac{K^{ab}\beta_{a}\beta_{b}}{\sqrt{1+\beta^{2}}}\nonumber \\
		&  & +\frac{1}{\sqrt{1+\beta^{2}}}\beta^{a}\dot{\beta}_{a}+a^{a}\beta_{a}+\mathrm{D}^{a}\beta_{a},
	\end{eqnarray}
which is the covariant 3+1 correspondence of $K$.
Clearly in the unitary gauge $n_{a} \rightarrow u_{a}$, i.e., in the limit $\beta_{a} \rightarrow 0$, the above reduces to $K$.
On the other hand, generally $\dot{\beta}_{a}$ arises, which signals the extra DOF's when deviating from the unitary gauge.

\section{Degenerate analysis: $d=2$} \label{sec:d2}

In the above we have derived the explicit covariant 3+1 correspondences of various SCG quantities.
When deviating from the unitary gauge, there arise extra Lie derivatives of $\beta_{a}$ and/or $K_{ab}$ (with coefficients proportional to $\beta_{a}$), which correspond to higher temporal derivatives of the scalar field and/or the metric.
This also explains the apparent appearance of extra modes for the SCG theory in general coordinates \cite{DeFelice:2018ewo,Iyonaga:2018vnu}.
It is possible, however, that such ``dangerous'' terms can get cancelled by combining several SCG terms.
In other words, there might exist particular SCG combinations, of which the COD formulation is also degenerate. Since the COD and GST are exactly equivalent, this means the corresponding GST is degenerate.

As a simple example, in this section we consider the linear combination
\begin{equation}
	\mathcal{L}^{\left(2\right)}_{\mathrm{SCG}}=c_{1}K_{ij}K^{ij}+c_{2}K^{2}+c_{3}\,{}^{3}\!R+c_{4}a_{i}a^{i},\label{calL_2}
\end{equation}
where the coefficients $c_{i}$'s are functions of $t$ and $N$.
The Lagrangian in (\ref{calL_2}) is the combination of 4 SCG monomials with $d=2$, where $d$ is the total number of the derivatives (temporal or spatial) in each monomial.
We refer to \cite{Gao:2020yzr} for more details on the classification of SCG monomials according to the derivatives.
The unitary gauge correspondence of (\ref{calL_2}) reads
\begin{equation}
	\mathcal{L}^{(2)}_{\mathrm{u.g.}}=c_{1}\,\overset{u}{K}_{ab}\overset{u}{K}{}^{ab}+c_{2}\,\overset{u}{K}{}^{2}+c_{3}\,{}^{3}\!\overset{u}{R}+c_{4}\,\overset{u}{a}_{a}\overset{u}{a}{}^{a}.\label{Lagd2}
\end{equation}
In (\ref{calL_2}), the coefficients $c_{i}$'s are understood as
functions of the scalar field $\phi$ as well as its canonical kinetic
term $X$.

In the spatially covariant formulation, only the spatial metric acquires
kinetic term through the extrinsic curvature. In the covariant correspondence,
extra terms carrying temporal derivative arise. In the current case,
these are $\dot{\beta}_{a}$ (i.e., $\dot{X}$) and $\dot{K}_{ab}$.
Therefore it is convenient to group terms according to the orders
of temporal derivatives of each term. After some manipulations, the
full covariant 3+1 correspondence can be written as
\begin{eqnarray}
	\mathcal{L}^{(2)}_{\mathrm{COD}} & = & \left.\mathcal{L}^{(2)}_{\mathrm{COD}}\right|_{\dot{\beta}^{2}}+\left.\mathcal{L}^{(2)}_{\mathrm{COD}}\right|_{\dot{\beta}K}+\left.\mathcal{L}^{(2)}_{\mathrm{COD}}\right|_{\dot{K}}+\left.\mathcal{L}^{(2)}_{\mathrm{COD}}\right|_{K^{2}}\nonumber \\
	&  & +\left.\mathcal{L}^{(2)}_{\mathrm{COD}}\right|_{\dot{\beta}}+\left.\mathcal{L}^{(2)}_{\mathrm{COD}}\right|_{K}+\left.\mathcal{L}^{(2)}_{\mathrm{COD}}\right|_{0}.
\end{eqnarray}

There are 4 kinds of terms that are of the second order in temporal
derivatives, which are
\begin{eqnarray}
	\left.\mathcal{L}^{(2)}_{\mathrm{COD}}\right|_{\dot{\beta}^{2}} & = & \dot{\beta}_{a}\dot{\beta}^{a}\left[c_{4}+\frac{1}{2}\left(c_{1}+c_{3}+2c_{4}\right)\beta^{2}\right]\nonumber \\
	&  & +\left(\dot{\beta}_{a}\beta^{a}\right)^{2}\left[-\frac{1}{2}\left(c_{1}+c_{3}+2c_{4}\right)+\frac{c_{1}+c_{2}}{1+\beta^{2}}\right],
\end{eqnarray}
\begin{eqnarray}
	\left.\mathcal{L}^{(2)}_{\mathrm{COD}}\right|_{\dot{\beta}\,K} & = & +2\left(c_{1}+c_{3}\right)\dot{\beta}_{a}\beta_{b}K^{ab}\nonumber \\
	&  & -\frac{2\left(c_{1}+c_{2}\right)}{1+\beta^{2}}\left(\dot{\beta}_{a}\beta^{a}\right)\left(K_{cd}\beta^{c}\beta^{d}\right)\nonumber \\
	&  & +2\left(c_{2}-c_{3}\right)\left(\dot{\beta}_{a}\beta^{a}\right)K,
\end{eqnarray}
\begin{equation}
	\left.\mathcal{L}^{(2)}_{\mathrm{COD}}\right|_{\dot{K}}=2c_{3}\left(\beta^{a}\beta^{b}-h^{ab}\beta^{2}\right)\dot{K}_{ab},
\end{equation}
and
\begin{eqnarray}
	\left.\mathcal{L}^{(2)}_{\mathrm{COD}}\right|_{K^{2}} & = & \left[c_{1}+(c_{1}+3c_{3})\beta^{2}\right]K_{ab}K^{ab}\nonumber \\
	&  & +\left[c_{2}+(c_{2}-c_{3})\beta^{2}\right]K^{2}\\
	&  & -2\left(c_{2}-2c_{3}\right)KK_{ab}\beta^{a}\beta^{b}\nonumber \\
	&  & -2\left(c_{1}+3c_{3}\right)K_{a}^{c}K_{bc}\beta^{a}\beta^{b}\nonumber \\
	&  & +\frac{c_{1}+c_{2}}{1+\beta^{2}}\left(K_{ab}\beta^{a}\beta^{b}\right)^{2}.
\end{eqnarray}
The terms of the first order in temporal derivatives are
\begin{eqnarray}
	\left.\mathcal{L}^{(2)}_{\mathrm{COD}}\right|_{\dot{\beta}} & = & +\frac{(c_{1}+c_{3})}{\sqrt{1+\beta^{2}}}(\dot{\beta}_{a}\beta_{b}\mathrm{D}^{a}\beta^{b})+(c_{1}+c_{3}+2c_{4})\sqrt{1+\beta^{2}}(\dot{\beta}_{a}\beta_{b}\mathrm{D}^{b}\beta^{a})\nonumber \\
	&  & +\frac{1}{\sqrt{1+\beta^{2}}}\left[2(c_{2}-c_{3})(\mathrm{D}_{c}\beta^{c})-(c_{1}+c_{3}+2c_{4})(\beta^{c}\beta^{d}\mathrm{D}_{c}\beta_{d})\right](\dot{\beta}_{a}\beta^{a})\nonumber \\
	&  & +\sqrt{1+\beta^{2}}\left[2c_{4}+(c_{1}+c_{3}+2c_{4})\beta^{2}\right](a^{a}\dot{\beta}_{a})\nonumber \\
	&  & +\frac{1}{\sqrt{1+\beta^{2}}}\left[c_{1}+2c_{2}-c_{3}-2c_{4}-(c_{1}+c_{3}+2c_{4})\beta^{2}\right](a^{c}\beta_{c})(\dot{\beta}_{a}\beta^{a}),
\end{eqnarray}
and
\begin{eqnarray}
	\left.\mathcal{L}^{(2)}_{\mathrm{COD}}\right|_{K} & = & -\frac{2(c_{2}-c_{3})}{\sqrt{1+\beta^{2}}}\left(K_{ab}\beta^{a}\beta^{b}\right)\left(\mathrm{D}_{c}\beta^{c}\right)-\frac{2(c_{1}+c_{3})}{\sqrt{1+\beta^{2}}}(K_{ab}\beta^{c}\beta^{a}\mathrm{D}^{b}\beta_{c})\nonumber \\
	&  & +2(c_{1}+c_{3})\sqrt{1+\beta^{2}}(a^{a}K_{ab}\beta^{b})-4c_{3}\sqrt{1+\beta^{2}}(\beta^{a}\mathrm{D}_{a}K)\nonumber \\
	&  & +4c_{3}\sqrt{1+\beta^{2}}(\beta^{a}\mathrm{D}_{b}K_{a}^{b})+2(c_{1}+c_{3})\sqrt{1+\beta^{2}}(K_{ab}\mathrm{D}^{b}\beta^{a})\nonumber \\
	&  & +2(c_{2}-c_{3})\sqrt{1+\beta^{2}}K(a^{a}\beta_{a})+2(c_{2}-c_{3})\sqrt{1+\beta^{2}}K(\mathrm{D}_{a}\beta^{a})\nonumber \\
	&  & -\frac{2(c_{1}+c_{2})}{\sqrt{1+\beta^{2}}}(a^{c}\beta_{c})(K_{ab}\beta^{a}\beta^{b}),
\end{eqnarray}
The terms containing no temporal derivative are
\begin{eqnarray}
	\left.\mathcal{L}^{(2)}_{\mathrm{COD}}\right|_{0} & = & c_{3}\,{}^{3}\!R+2c_{3}(\,{}^{3}\!R_{ab}\beta^{a}\beta^{b})+(c_{2}-c_{3})(\mathrm{D}_{a}\beta^{a})^{2}+(c_{1}+c_{3})(a^{a}\beta^{b}\mathrm{D}_{a}\beta_{b})+2(\mathrm{D}_{a}a^{a})c_{3}\beta^{2}\nonumber \\
	&  & -2c_{3}(\beta^{a}\beta^{b}\mathrm{D}_{b}a_{a})+(a^{a}\beta_{a})\left[2(c_{2}-c_{3})(\mathrm{D}_{c}\beta^{c})-(c_{1}+c_{3}+2c_{4})(\beta^{c}\beta^{d}\mathrm{D}_{c}\beta_{d})\right]\nonumber \\
	&  & +\frac{1}{2}(c_{1}+c_{3}+2c_{4})(\beta^{a}\beta^{c}\mathrm{D}_{a}\beta^{b}\mathrm{D}_{c}\beta_{b})+\frac{1}{2}(c_{1}+c_{3})(\mathrm{D}_{a}\beta_{b}\mathrm{D}^{b}\beta^{a})\nonumber \\
	&  & +\frac{1}{2}(c_{1}+c_{3})(\mathrm{D}_{b}\beta_{a}\mathrm{D}^{b}\beta^{a})+(c_{1}+c_{3}+2c_{4})(1+\beta^{2})(a^{a}\beta^{b}\mathrm{D}_{b}\beta_{a})\nonumber \\
	&  & -\frac{c_{1}+c_{3}+2c_{4}}{2(1+\beta^{2})}(\beta^{a}\beta^{b}\mathrm{D}_{b}\beta_{a})^{2}-\frac{c_{1}+c_{3}}{2(1+\beta^{2})}(\beta^{a}\beta^{c}\mathrm{D}_{b}\beta_{c}\mathrm{D}^{b}\beta_{a})\nonumber \\
	&  & +\frac{1}{2}(a^{a}\beta_{a})^{2}\left[c_{1}+2c_{2}-5c_{3}-2c_{4}-(c_{1}+c_{3}+2c_{4})\beta^{2}\right]\nonumber \\
	&  & +\frac{1}{2}(a_{a}a^{a})\left[2c_{4}+(c_{1}+5c_{3}+4c_{4})\beta^{2}+(c_{1}+c_{3}+2c_{4})\left(\beta^{2}\right)^{2}\right].
\end{eqnarray}

The presence of $\dot{\beta}^{2}$, $\dot{\beta}K$ and $\dot{K}$
terms correspond to the higher temporal derivatives, and thus signal the possible propagation of extra mode(s). Our goal is thus to tune the coefficients $c_1,\cdots,c_{4}$ such that all these ``dangerous'' terms are suppressed.
In the following, we replace $\beta_{a}$ (and its spatial derivatives) in terms
of the scalar field $\phi$, its kinetic term $X$ and their temporal and
spatial derivatives.

In the rest part of this work, we suppress the subscript ``COD'' for simplicity.
Schematically we write
\begin{eqnarray}
	\mathcal{L}^{(2)} & = & \left.\mathcal{L}^{(2)}\right|_{\dot{X}^{2}}+\left.\mathcal{L}^{(2)}\right|_{\dot{X}K}+\left.\mathcal{L}^{(2)}\right|_{\dot{K}}+\left.\mathcal{L}^{(2)}\right|_{K^{2}}\nonumber \\
	&  & +\left.\mathcal{L}^{(2)}\right|_{\dot{X}}+\left.\mathcal{L}^{(2)}\right|_{K}+\left.\mathcal{L}^{(2)}\right|_{0},
\end{eqnarray}
where the first line are monomials of the second order in the Lie derivative,
the second line are monomials of the first order in the Lie derivative and containing spatial derivatives only. For the terms of the second order in the Lie derivatives, we have
\begin{equation}
	\left.\mathcal{L}^{(2)}\right|_{\dot{X}^{2}}=\frac{\dot{X}^{2}\left(\mathrm{D}\phi\right)^{2}}{8X^{3}\dot{\phi}^{2}}\left[\left(c_{1}+c_{2}\right)\left(\mathrm{D}\phi\right)^{2}+c_{4}\dot{\phi}^{2}\right],\label{Lag_d2_X2}
\end{equation}
\begin{equation}
	\left.\mathcal{L}^{(2)}\right|_{\dot{X}K}=-\frac{\dot{X}}{2X^{2}\dot{\phi}^{2}}\left\{ K_{ab}\mathrm{D}^{a}\phi\mathrm{D}^{b}\phi\left[\left(c_{1}+c_{3}\right)\dot{\phi}^{2}-\left(c_{1}+c_{2}\right)\left(\mathrm{D}\phi\right)^{2}\right]+\left(c_{2}-c_{3}\right)K\left(\mathrm{D}\phi\right)^{2}\dot{\phi}^{2}\right\} ,
\end{equation}
\begin{equation}
	\left.\mathcal{L}^{(2)}\right|_{\dot{K}}=\frac{c_{3}}{X}\dot{K}_{ab}\left(\mathrm{D}^{a}\phi\mathrm{D}^{b}\phi-h^{ab}\left(\mathrm{D}\phi\right)^{2}\right),
\end{equation}
and
\begin{eqnarray}
	\left.\mathcal{L}^{(2)}\right|_{K^{2}} & = & -\frac{1}{2X\dot{\phi}^{2}}\Big[-\left(c_{2}\dot{\phi}^{2}-c_{3}\left(\mathrm{D}\phi\right)^{2}\right)K^{2}\dot{\phi}^{2}\nonumber \\
	&  & -\left(c_{1}\dot{\phi}^{2}+3c_{3}\left(\mathrm{D}\phi\right)^{2}\right)K_{ab}K^{ab}\dot{\phi}^{2}\nonumber \\
	&  & +2\left(c_{2}-2c_{3}\right)KK_{ab}\mathrm{D}^{a}\phi\mathrm{D}^{b}\phi\dot{\phi}^{2}\nonumber \\
	&  & +2\left(c_{1}+3c_{3}\right)K_{a}^{c}K_{bc}\mathrm{D}^{a}\phi\mathrm{D}^{b}\phi\dot{\phi}^{2}\nonumber \\
	&  & -\left(c_{1}+c_{2}\right)\left(K_{ab}\mathrm{D}^{a}\phi\mathrm{D}^{b}\phi\right)^{2}\Big].
\end{eqnarray}
We shall pay special attention to the terms involving $\dot{K}$,
which should be reduced by the integrations by parts using
\begin{equation}
	\mathcal{C}^{ab}\dot{K}_{ab}\simeq-K\mathcal{C}^{ab}K_{ab}-\left(\pounds_{\bm{n}}{\mathcal{C}}^{ab}\right)K_{ab}.
\end{equation}
After performing the integrations by parts, since the $\dot{K}$ terms
have been reduced, there are 3 types of terms that are second order in the Lie
derivatives. The $\dot{X}^{2}$ terms are not affected as in (\ref{Lag_d2_X2}),
while the $\dot{X}K$ and $K^{2}$ terms become
\begin{eqnarray}
	\left.\mathcal{L}^{(2)}\right|_{\dot{X}K} & = & -\frac{1}{2X^{2}\dot{\phi}^{2}}\dot{X}\Big[\left(c_{1}-c_{3}+2X\frac{\partial c_{3}}{\partial X}\right)K_{ab}\mathrm{D}^{a}\phi\mathrm{D}^{b}\phi\dot{\phi}^{2}\nonumber \\
	&  & \quad-\left(c_{1}+c_{2}\right)K_{ab}\mathrm{D}^{a}\phi\mathrm{D}^{b}\phi\left(\mathrm{D}\phi\right)^{2}+\left(c_{2}+c_{3}-2X\frac{\partial c_{3}}{\partial X}\right)K\left(\mathrm{D}\phi\right)^{2}\dot{\phi}^{2}\Big], \label{Lag_d2_dotXK}
\end{eqnarray}
and
\begin{eqnarray}
	\left.\mathcal{L}^{(2)}\right|_{K^{2}} & = & \frac{1}{2X\dot{\phi}^{2}}\Big[\left(c_{2}\dot{\phi}^{2}+c_{3}\left(\mathrm{D}\phi\right)^{2}\right)K^{2}\dot{\phi}^{2}+\left(c_{1}\dot{\phi}^{2}-c_{3}\left(\mathrm{D}\phi\right)^{2}\right)K_{ab}K^{ab}\dot{\phi}^{2}\nonumber \\
	&  & -2\left(c_{2}+c_{3}\right)KK_{ab}\mathrm{D}^{a}\phi\mathrm{D}^{b}\phi\dot{\phi}^{2}-2\left(c_{1}-c_{3}\right)K_{a}^{c}K_{bc}\mathrm{D}^{a}\phi\mathrm{D}^{b}\phi\dot{\phi}^{2}\nonumber \\
	&  & +\left(c_{1}+c_{2}\right)\left(K_{ab}\mathrm{D}^{a}\phi\mathrm{D}^{b}\phi\right)^{2}\Big].
\end{eqnarray}

We are now ready to determine the coefficients in order make the COD Lagrangian degenerate. 
\begin{enumerate}
	\item No $\dot{X}^{2}$ term: From (\ref{Lag_d2_X2}) we must set
	\begin{eqnarray}
		c_{1}+c_{2} & = & 0,\\
		c_{4} & = & 0.
	\end{eqnarray}
	\item No $\dot{X}K$ terms: From (\ref{Lag_d2_dotXK}) we must set
	\begin{eqnarray}
		c_{1}-c_{3}+2X\frac{\partial c_{3}}{\partial X} & = & 0,\\
		c_{2}+c_{3}-2X\frac{\partial c_{3}}{\partial X} & = & 0.
	\end{eqnarray}
\end{enumerate}
We have the unique solutions for the coefficients:
\begin{equation}
	c_{1}=-c_{2}=c_{3}-2X\frac{\partial c_{3}}{\partial X},\qquad c_{4}=0.\label{coeff_sol_d2}
\end{equation}
This is nothing but corresponds to the Horndeski Lagrangian in the unitary
gauge \cite{Gleyzes:2013ooa}.
In other words, the specific combination
	\begin{equation}
		\mathcal{L}^{\left(2\right)}_{\mathrm{SCG}}=\left(c_{3}-2X\frac{\partial c_{3}}{\partial X}\right)\left(K_{ij}K^{ij}-K^{2}\right)+c_{3}\,{}^{3}\!R,
	\end{equation}
represents the SCG Lagrangian of which the corresponding GST is degenerate\footnote{The corresponding GST is the Horndeski Lagrangian $\mathcal{L}_{4}$ (in the convention of \cite{Deffayet:2011gz,Kobayashi:2011nu}).}.
Clearly the GR is a special case with $c_3$ being constant.

It is interesting to check, after applying the degeneracy conditions, 
\begin{eqnarray}
	\left.\mathcal{L}^{(2)}\right|_{K^{2}} & \rightarrow & \left(c_{3}-2X\frac{\partial c_{3}}{\partial X}\right)\left(K_{ab}K^{ab}-K^{2}\right)\nonumber \\
	&  & +\frac{\partial c_{3}}{\partial X}\left(-2h^{cd}\mathrm{D}^{a}\phi\mathrm{D}^{b}\phi+2h^{bc}\mathrm{D}^{a}\phi\mathrm{D}^{d}\phi-\left(h^{ac}h^{bd}-h^{ab}h^{cd}\right)\left(\mathrm{D}\phi\right)^{2}\right)K_{cd}K_{ab}.
\end{eqnarray}
The second line is proportional to $\mathrm{D}_{a}\phi$ and thus is vanishing in the unitary gauge.

After imposing the above conditions, at the linear order in the Lie derivatives, there are terms proportional to $K$ and $\dot{X}$. 
For terms proportional to $\dot{X}$, we find
	\begin{equation}
		\left.\mathcal{L}^{(2)}\right|_{\dot{X}}=\frac{\dot{X}}{X^{2}\dot{\phi}}\left(c_{3}-X\frac{\partial c_{3}}{\partial X}\right)\left(\mathrm{D}^{a}\phi\mathrm{D}_{a}\mathrm{D}_{b}\phi\mathrm{D}^{b}\phi-\mathrm{D}^{2}\phi\left(\mathrm{D}\phi\right)^{2}\right).
	\end{equation}
These two types of terms are safe since they have nothing to do with the degeneracy, which can also be further reduced by the integrations by parts.

\section{Degenerate analysis: $d=3$ without $a_{i}$} \label{sec:d3}

In this section, we consider the SCG Lagrangian 
\begin{eqnarray}
	\mathcal{L}^{\left(3\right)} & = & c_{1}^{\left(0;3,0\right)}K_{ij}K^{jk}K_{k}^{i}+c_{2}^{\left(0;3,0\right)}K_{ij}a^{i}a^{j}+c_{3}^{\left(0;3,0\right)}K_{ij}K^{ij}K+c_{4}^{\left(0;3,0\right)}Ka_{i}a^{i}+c_{5}^{\left(0;3,0\right)}K^{3}\nonumber \\
	&  & +c_{1}^{\left(0;1,1\right)}K_{ij}\nabla^{i}a^{j}+c_{2}^{\left(0;1,1\right)}K\nabla_{i}a^{i}\nonumber \\
	&  & +c_{1}^{\left(1;1,0\right)}\,{}^{3}\!R^{ij}K_{ij}+c_{2}^{\left(1;1,0\right)}\,{}^{3}\!RK,\label{calL_3}
\end{eqnarray}
which is the linear combination of SCG monomials of $d=3$.
In (\ref{calL_3}) all the coefficients $c_{n}^{\left(c_0;d_2,d_3\right)}$ are functions of $t$ and $N$.
We refer to \cite{Gao:2020yzr} for details on the meaning of the superscripts.
In this section, we turn off the terms involving the acceleration
$a_{i}$, i.e., we set
\begin{equation}
	c_{2}^{\left(0;3,0\right)}=c_{4}^{\left(0;3,0\right)}=c_{1}^{\left(0;1,1\right)}=c_{2}^{\left(0;1,1\right)}=0.
\end{equation}

\subsection{The third order in the Lie derivative}

At the third order in the Lie derivatives, schematically, there are
in total 6 types of monomials, of which 5 are dangerous:
\begin{equation}
	\dot{X}^{3},\quad\dot{X}^{2}K,\quad\dot{X}\dot{K},\quad K\dot{K},\quad\dot{X}K^{2}, \label{terms_d3_3a}
\end{equation}
and 1 is safe:
	\begin{equation}
		K^{3}. \label{terms_d3_3b}
	\end{equation}

At the third order in the Lie derivatives, $\dot{X}^{3},\dot{X}^{2}K,\dot{X}\dot{K}$
terms cannot be reduced by integrations by parts\footnote{Although the $\dot{X}\dot{K}$ term can also be transformed by the integration by parts:
$\mathcal{F}\dot{X}\dot{K}\simeq-K\mathcal{F}\dot{X}K-\dot{\mathcal{F}}\dot{X}K-\mathcal{F}\ddot{X}K$, we find it is not necessary since the new term $\ddot{X}K$ will arise. Therefore we simply keep the $\dot{X}\dot{K}$ term in its original form.}. Therefore we must to suppress them by setting the corresponding coefficients to be vanishing identically.
On the other hand, the terms involving $\dot{K}$ should be reduced by the integrations by parts. 
For the $K\dot{K}$ term, schematically we write
	\begin{equation}
	\mathcal{C}^{ab,cd}K_{cd}\dot{K}_{ab}\simeq-\frac{1}{2}K\mathcal{C}^{ab,cd}K_{cd}K_{ab}-\frac{1}{2}\left(\pounds_{\bm{n}}\mathcal{C}^{ab,cd}\right)K_{ab}K_{cd}+\frac{1}{2}\left(\mathcal{C}^{ab,cd}-\mathcal{C}^{cd,ab}\right)K_{cd}\dot{K}_{ab},\label{ibp_KdK}
	\end{equation}
which cannot be reduced further. After performing the integration
by parts, the $K\dot{K}$ terms should eliminated by tuning the coefficients,
if not being vanishing identically.

After performing the integration by parts of $K\dot{K}$ terms using
(\ref{ibp_KdK}), for the $\dot{X}^{3}$ terms, we find 
\begin{equation}
	\left.\mathcal{L}_{3}^{\left(3\right)}\right|_{\dot{X}^{3}}=-\frac{\dot{X}^{3}\left(\mathrm{D}\phi\right)^{6}}{\left(2X\right)^{9/2}\dot{\phi}^{3}}\left(c_{1}^{(0;3,0)}+c_{3}^{(0;3,0)}+c_{5}^{(0;3,0)}\right),
\end{equation}
therefore we need to impose one condition:
\begin{equation}
	c_{1}^{(0;3,0)}+c_{3}^{(0;3,0)}+c_{5}^{(0;3,0)}=0,\label{cond_d3_1}
\end{equation}
from which we solve
\begin{equation}
	c_{5}^{(0;3,0)}=-c_{1}^{(0;3,0)}-c_{3}^{(0;3,0)}.\label{coeff_d3_1}
\end{equation}

For the $\dot{X}^{2}K$ terms, we have
	\begin{eqnarray}
		\left.\mathcal{L}_{3}^{\left(3\right)}\right|_{\dot{X}^{2}K} & = & \frac{\dot{X}^{2}\left(\mathrm{D}\phi\right)^{2}}{\left(2X\right)^{7/2}\dot{\phi}^{3}}\Big[2XK_{ab}\mathrm{D}^{a}\phi\mathrm{D}^{b}\phi\left(3c_{1}^{(0;3,0)}+c_{1}^{(1;1,0)}+2c_{2}^{(1;1,0)}+2c_{3}^{(0;3,0)}\right)\nonumber \\
		&  & \quad+\left(K\left(\mathrm{D}\phi\right)^{2}\dot{\phi}^{2}-\left(\mathrm{D}\phi\right)^{2}K_{ab}\mathrm{D}^{a}\phi\mathrm{D}^{b}\phi\right)\left(-c_{1}^{(1;1,0)}-2c_{2}^{(1;1,0)}+c_{3}^{(0;3,0)}+3c_{5}^{(0;3,0)}\right)\Big],
	\end{eqnarray}
After applying the condition (\ref{cond_d3_1}), the above is reduced to be
	\begin{equation}
		\left.\mathcal{L}_{3}^{\left(3\right)}\right|_{\dot{X}^{2}K}\rightarrow-\frac{\dot{X}^{2}\left(\mathrm{D}\phi\right)^{2}}{\left(2X\right)^{7/2}\dot{\phi}}\left(K\left(\mathrm{D}\phi\right)^{2}-K_{ab}\mathrm{D}^{a}\phi\mathrm{D}^{b}\phi\right)\left(3c_{1}^{(0;3,0)}+c_{1}^{(1;1,0)}+2c_{2}^{(1;1,0)}+2c_{3}^{(0;3,0)}\right).
	\end{equation}
Thus we need to impose the second condition
\begin{equation}
	3c_{1}^{(0;3,0)}+c_{1}^{(1;1,0)}+2c_{2}^{(1;1,0)}+2c_{3}^{(0;3,0)}=0,\label{cond_d3_2}
\end{equation}
from which we solve
\begin{equation}
	c_{3}^{(0;3,0)}=-\frac{3}{2}c_{1}^{(0;3,0)}-\frac{1}{2}c_{1}^{(1;1,0)}-c_{2}^{(1;1,0)}.\label{coeff_d3_2}
\end{equation}

For the $\dot{X}\dot{K}$ terms, we find
	\begin{equation}
		\left.\mathcal{L}_{3}^{\left(3\right)}\right|_{\dot{X}\dot{K}}=\frac{\dot{X}\left(\mathrm{D}\phi\right)^{2}}{\left(2X\right)^{5/2}\dot{\phi}}\dot{K}_{ab}\left(h^{ab}\left(\mathrm{D}\phi\right)^{2}-\mathrm{D}^{a}\phi\mathrm{D}^{b}\phi\right)\left(c_{1}^{(1;1,0)}+2c_{2}^{(1;1,0)}\right). \label{Lag_d3_dotXdotK}
	\end{equation}
In deriving (\ref{Lag_d3_dotXdotK}) we have not used the conditions (\ref{cond_d3_1}) and (\ref{cond_d3_2}).
Therefore we need to impose the third condition
\begin{equation}
	c_{1}^{(1;1,0)}+2c_{2}^{(1;1,0)}=0,\label{cond_d3_3}
\end{equation}
from which we solve
\begin{equation}
	c_{2}^{(1;1,0)}=-\frac{1}{2}c_{1}^{(1;1,0)}.\label{coeff_d3_3}
\end{equation}

Using (\ref{coeff_d3_3}), (\ref{coeff_d3_2}) is reduced to be
\begin{equation}
	c_{3}^{(0;3,0)}=-\frac{3}{2}c_{1}^{(0;3,0)}.\label{coeff_d3_2a}
\end{equation}
Plugging (\ref{coeff_d3_2a}) into (\ref{coeff_d3_1}) yields
\begin{equation}
	c_{5}^{(0;3,0)}=\frac{1}{2}c_{1}^{(0;3,0)}.\label{coeff_d3_1a}
\end{equation}

For the $K\dot{K}$ terms, we find
	\begin{equation}
		\left.\mathcal{L}_{3}^{\left(3\right)}\right|_{K\dot{K}}=-\frac{c_{1}^{(1;1,0)}+2c_{2}^{(1;1,0)}}{2\sqrt{2X}\dot{\phi}}\dot{K}_{ab}K_{cd}\left(h^{ab}\mathrm{D}^{c}\phi\mathrm{D}^{d}\phi-h^{cd}\mathrm{D}^{a}\phi\mathrm{D}^{b}\phi\right).
	\end{equation}
Fortunately, this term gets cancelled exactly after imposing
the condition (\ref{coeff_d3_3}). Therefore, after performing the
integration by parts and imposing the condition (\ref{coeff_d3_3}),
the $K\dot{K}$ terms are removed automatically.

Then we are left with only the $\dot{X}K^{2}$ terms, which have two
origins. One corresponds to those already exist in the original
expression, the other corresponds to those arise from $K\dot{K}$
terms after the integration by parts. The full expression of $\dot{X}K^{2}$
terms without the above degeneracy conditions are tedious, which we
do not present in the current work. After applying all the above three conditions (\ref{cond_d3_1}), (\ref{cond_d3_2}) and (\ref{cond_d3_3}), we find that
	\begin{eqnarray}
		\left.\mathcal{L}_{3}^{\left(3\right)}\right|_{\dot{X}K^{2}} & = & -\frac{\dot{X}\dot{\phi}}{2\left(2X\right)^{5/2}}\left(3c_{1}^{(0;3,0)}+2X\frac{\partial c_{1}^{(1;1,0)}}{\partial X}\right)\nonumber \\
		&  & \times\left[\left(K^{2}-K_{ab}K^{ab}\right)\left(\mathrm{D}\phi\right)^{2}-2KK_{ab}\mathrm{D}^{a}\phi\mathrm{D}^{b}\phi+2K_{a}^{c}K_{bc}\mathrm{D}^{a}\phi\mathrm{D}^{b}\phi\right].
	\end{eqnarray}
In order to remove this term, we need to impose the fourth condition
\begin{equation}
	3c_{1}^{(0;3,0)}+2X\frac{\partial c_{1}^{(1;1,0)}}{\partial X}=0,
\end{equation}
from which we solve
\begin{equation}
	c_{1}^{(0;3,0)}=-\frac{2}{3}X\frac{\partial c_{1}^{(1;1,0)}}{\partial X}.\label{coeff_d3_4}
\end{equation}

It is interesting that, at the third order in the Lie derivatives, we have already got the whole 4 conditions (\ref{coeff_d3_1a}), (\ref{coeff_d3_2a}), (\ref{coeff_d3_3}) and (\ref{coeff_d3_4}) in the Horndeski theory of $\mathcal{L}_{5}$ \cite{Gleyzes:2013ooa}.

\subsection{The second and the first orders in Lie derivatives}

As a consistency check, in the following we shall show
that all the dangerous terms at the second and first order in Lie
derivatives are indeed removed.

At the second order in the Lie derivatives, schematically, there are in total 4 types of monomials, of which 3 are dangerous:
	\begin{equation}
		\dot{X}^{2},\quad\dot{X}\,K,\quad\dot{K},
	\end{equation}
and 1 is safe:
	\begin{equation}
		K^{2}.
	\end{equation}
The terms involving $\dot{K}$ can be fully reduced by using
\begin{equation}
	\mathcal{C}^{ab}\dot{K}_{ab}\simeq-K\mathcal{C}^{ab}K_{ab}- \left(\pounds_{\bm{n}}\mathcal{C}^{ab} \right)K_{ab},
\end{equation}
where $\mathcal{C}^{ab}$ contains no Lie derivative.
For the terms of the second order in the Lie derivatives, after imposing the 4 conditions (\ref{coeff_d3_1a}), (\ref{coeff_d3_2a}), (\ref{coeff_d3_3}) and (\ref{coeff_d3_4}), we have examined that all the ``dangerous'' terms (i.e., involving $\dot{X}^2$, $\dot{K}$ and $\dot{X}K$) get cancelled automatically. Therefore we do not need to impose any further condition.

There are two types of terms of the first order in Lie derivatives, $\dot{X}$ and $K$. These two types of terms are always safe. Nevertheless, it is interesting to see that after imposing the 4 conditions (\ref{coeff_d3_1a}), (\ref{coeff_d3_2a}),
(\ref{coeff_d3_3}) and (\ref{coeff_d3_4}), 
	\begin{eqnarray}
		\left.\mathcal{L}_{1}^{(3)}\right|_{\dot{X}} & = & \frac{\dot{X}}{2\left(2X\right)^{5/2}\dot{\phi}}\bigg\{-c_{1}^{(1;1,0)}4XG_{ab}\mathrm{D}^{a}\phi\mathrm{D}^{b}\phi\nonumber \\
		&  & \quad+\left(3c_{1}^{(1;1,0)}-2X\frac{\partial c_{1}^{(1;1,0)}}{\partial X}\right)\Big[\left(\mathrm{D}\phi\right)^{2}\mathrm{D}_{a}\mathrm{D}_{b}\phi\mathrm{D}^{a}\mathrm{D}^{b}\phi-\left(\mathrm{D}\phi\right)^{2}\left(\mathrm{D}^{2}\phi\right)^{2}\nonumber \\
		&  & \qquad+2\left(\mathrm{D}^{2}\phi\right)\mathrm{D}^{a}\phi\mathrm{D}^{b}\phi\mathrm{D}_{a}\mathrm{D}_{b}\phi-2\mathrm{D}^{a}\phi\mathrm{D}^{b}\phi\mathrm{D}_{a}\mathrm{D}_{c}\phi\mathrm{D}^{c}\mathrm{D}_{b}\phi\Big]\bigg\}.
	\end{eqnarray}
Moreover, after the integrations by parts, there also arise terms involving
the Lie derivatives of the acceleration $\dot{a}_{a}$, which are possibly dangerous. We have checked that these terms are exactly cancelled out after imposing the 4 conditions (\ref{coeff_d3_1a}), (\ref{coeff_d3_2a}),
(\ref{coeff_d3_3}) and (\ref{coeff_d3_4}).

\section{Degenerate analysis: $d=3$ with $a_{i}$} \label{sec:d3a}

In this section, we consider the Lagrangian of $d=3$ (\ref{calL_3}) with all the coefficients are present.
As in the previous section, we first focus on the terms of the third order in Lie
derivatives. Due to the presence of $\nabla_{i}a_{j}$ terms, there
arise $\ddot{X}$ terms. Schematically, there are two types of terms
\begin{equation}
\ddot{X}\dot{X},\qquad\ddot{X}K,
\end{equation}
due to the presence of $a_{i}$. Nevertheless, by performing
the integrations by parts
\begin{equation}
\mathcal{C}\ddot{X}\dot{X}\simeq-\frac{1}{2}\left(K\mathcal{C}+\dot{\mathcal{C}}\right)\dot{X}^{2},
\end{equation}
and
\begin{equation}
\mathcal{C}^{ab}\ddot{X}K_{ab}\simeq-K\mathcal{C}^{ab}\dot{X}K_{ab}-\left(\pounds_{\bm{n}}\mathcal{C}^{ab}\right)\dot{X}K_{ab}-\mathcal{C}^{ab}\dot{X}\dot{K}_{ab},
\end{equation}
the two terms $\ddot{X}\dot{X}$ and $\ddot{X}K$ can be reduced to the 6 types of terms in (\ref{terms_d3_3a}) and (\ref{terms_d3_3b}) that already exist in the case without $a_{i}$.

In the following, we first perform the integrations by parts to reduce
the $\ddot{X}$ terms, then make a similar analysis as in Sec. \ref{sec:d3}. For
the $\dot{X}^{3}$ terms, we have
\begin{eqnarray}
\left.\mathcal{L}_{3}^{(3)}\right|_{\dot{X}^{3}} & = & -\frac{\dot{X}^{3}\left(\mathrm{D}\phi\right)^{4}}{2\left(2X\right)^{9/2}\dot{\phi}^{3}}\bigg[2X\dot{\phi}^{2}\frac{\partial\left(c_{1}^{(0;1,1)}+c_{2}^{(0;1,1)}\right)}{\partial X}-4X\left(c_{1}^{(0;1,1)}+c_{2}^{(0;1,1)}-c_{2}^{(0;3,0)}-c_{4}^{(0;3,0)}\right)\nonumber \\
&  & +\left(-c_{1}^{(0;1,1)}-c_{2}^{(0;1,1)}+2c_{1}^{(0;3,0)}+2c_{2}^{(0;3,0)}+2c_{3}^{(0;3,0)}+2c_{4}^{(0;3,0)}+2c_{5}^{(0;3,0)}\right)\left(\mathrm{D}\phi\right)^{2}\bigg].
\end{eqnarray}
We must set
\begin{eqnarray}
\frac{\partial\left(c_{1}^{(0;1,1)}+c_{2}^{(0;1,1)}\right)}{\partial X} & = & 0,\\
c_{1}^{(0;1,1)}+c_{2}^{(0;1,1)}-c_{2}^{(0;3,0)}-c_{4}^{(0;3,0)} & = & 0,\\
-c_{1}^{(0;1,1)}-c_{2}^{(0;1,1)}+2c_{1}^{(0;3,0)}+2c_{2}^{(0;3,0)}+2c_{3}^{(0;3,0)}+2c_{4}^{(0;3,0)}+2c_{5}^{(0;3,0)} & = & 0,
\end{eqnarray}
We solve
\begin{eqnarray}
c_{4}^{(0;3,0)} & = & -c_{2}^{(0;3,0)}+f_{1}\left(\phi\right),\label{coeff_d3a_11}\\
c_{2}^{(0;1,1)} & = & -c_{1}^{(0;1,1)}+f_{1}\left(\phi\right),\label{coeff_d3a_12}\\
c_{5}^{(0;3,0)} & = & -c_{1}^{(0;3,0)}-c_{3}^{(0;3,0)}-\frac{1}{2}f_{1}\left(\phi\right),\label{coeff_d3a_13}
\end{eqnarray}
where $f_{1}(\phi)$ is an arbitrary function of $\phi$ only.

For the $\dot{X}^{2}K$ terms, after applying the above conditions
(\ref{coeff_d3a_11})-(\ref{coeff_d3a_13}), we have
\begin{eqnarray}
\left.\mathcal{L}_{3}^{(3)}\right|_{\dot{X}^{2}K} & = & -\frac{\dot{X}^{2}}{\left(2X\right)^{7/2}\dot{\phi}}\left(K\left(\mathrm{D}\phi\right)^{2}-K_{ab}\mathrm{D}^{a}\phi\mathrm{D}^{b}\phi\right)\nonumber \\
&  & \times\bigg[\left(-c_{1}^{(0;1,1)}+3c_{1}^{(0;3,0)}+c_{1}^{(1;1,0)}+c_{2}^{(0;3,0)}+2c_{2}^{(1;1,0)}+2c_{3}^{(0;3,0)}+2f_{1}(\phi)\right)\left(\mathrm{D}\phi\right)^{2}\nonumber \\
&  & \quad+2X\dot{\phi}^{2}\frac{\partial c_{1}^{(0;1,1)}}{\partial X}+X\left(-2c_{1}^{(0;1,1)}+2c_{2}^{(0;3,0)}\right)\bigg].
\end{eqnarray}
We must have
\begin{eqnarray}
-c_{1}^{(0;1,1)}+3c_{1}^{(0;3,0)}+c_{1}^{(1;1,0)}+c_{2}^{(0;3,0)}+2c_{2}^{(1;1,0)}+2c_{3}^{(0;3,0)}+2f_{1}(\phi) & = & 0,\\
\frac{\partial c_{1}^{(0;1,1)}}{\partial X} & = & 0,\\
-2c_{1}^{(0;1,1)}+2c_{2}^{(0;3,0)} & = & 0,
\end{eqnarray}
from which we solve
\begin{eqnarray}
c_{1}^{(0;1,1)} & = & f_{2}\left(\phi\right),\label{coeff_d3a_21}\\
c_{2}^{(0;3,0)} & = & f_{2}\left(\phi\right),\label{coeff_d3a_22}\\
c_{3}^{(0;3,0)} & = & -\frac{3c_{1}^{(0;3,0)}}{2}-\frac{c_{1}^{(1;1,0)}}{2}-c_{2}^{(1;1,0)}-f_{1}\left(\phi\right).\label{coeff_d3a_23}
\end{eqnarray}
Again, $f_{2}(\phi)$ is an arbitrary function of $\phi$ only.

For the $\dot{X}\dot{K}$ terms, after applying the above conditions
(\ref{coeff_d3a_11})-(\ref{coeff_d3a_13}) and (\ref{coeff_d3a_21})-(\ref{coeff_d3a_23}),
we have
\begin{eqnarray}
\left.\mathcal{L}_{3}^{(3)}\right|_{\dot{X}\dot{K}} & = & \frac{\dot{X}}{\left(2X\right)^{5/2}\dot{\phi}}\bigg[\left(c_{1}^{(1;1,0)}+2c_{2}^{(1;1,0)}+f_{1}(\phi)-f_{2}(\phi)\right)(h^{ab}\dot{K}_{ab})\left(\mathrm{D}\phi\right)^{4}\nonumber \\
&  & \quad-\left(c_{1}^{(1;1,0)}+2c_{2}^{(1;1,0)}+f_{1}(\phi)-f_{2}(\phi)\right)\left(\mathrm{D}\phi\right)^{2}\dot{K}_{ab}\mathrm{D}^{a}\phi\mathrm{D}^{b}\phi\nonumber \\
&  & \quad+2\left(f_{1}(\phi)-f_{2}(\phi)\right)X(h^{ab}\dot{K}_{ab})\left(\mathrm{D}\phi\right)^{2}+2f_{2}(\phi)X\dot{K}_{ab}\mathrm{D}^{a}\phi\mathrm{D}^{b}\phi\bigg].
\end{eqnarray}
We must have
\begin{eqnarray}
c_{1}^{(1;1,0)}+2c_{2}^{(1;1,0)}+f_{1}(\phi)-f_{2}(\phi) & = & 0,\\
f_{1}(\phi)-f_{2}(\phi) & = & 0,\\
f_{2}(\phi) & = & 0,
\end{eqnarray}
from which we solve
\begin{equation}
f_{1}\left(\phi\right)=f_{2}\left(\phi\right)=0,
\end{equation}
and
\begin{equation}
c_{1}^{(1;1,0)}+2c_{2}^{(1;1,0)}=0.
\end{equation}

At this point, we can already fix that
\begin{equation}
c_{2}^{\left(0;3,0\right)}=c_{4}^{\left(0;3,0\right)}=c_{1}^{\left(0;1,1\right)}=c_{2}^{\left(0;1,1\right)}=0,
\end{equation}
and therefore we have been back to the case without $a_{i}$.
As a result, the subsequent analysis is exactly the same as the case without $a_{i}$ in Sec. \ref{sec:d3}.

\section{Conclusion} \label{sec:con}

A necessary condition for a generally covariant scalar-tensor theory (GST) to be ghostfree is that it is ghostfree in the unitary gauge when the scalar field is timelike, in which the theory takes the form of the spatially covariant gravity (SCG).
One may use the SCG as the starting point to search for the ghostfree GST.
To this end, a further covariant 3+1 decomposition (COD) of the GST without fixing any coordinates is also needed.
Therefore in principle one needs ``two steps'' (SCG$\rightarrow $GST$\rightarrow $COD) to complete the analysis.
In this work, we developed a ``one step'' method, which we dub the ``covariant 3+1 correspondence'', to derive the corresponding COD from SCG directly.
The resulting COD expressions can be used as the starting point of the further degeneracy/constraint analysis.

In Sec. \ref{sec:covcorr} we derive the explicit expressions of this covariant 3+1 correspondence.
We take the SCG Lagrangians of $d=2$ and $d=3$ as simple illustrations of this method in the subsequent sections.
By deriving the corresponding COD using this method, one can determine the degeneracy conditions easily.
No surprisingly, the resulting Lagrangians with these degenracy conditions are nothing but correspond to the Horndeski theory in the unitary gauge.
In other words, one could re-discover the Horndeski theory with this method in a quite simple manner.
In this work, we only consider SCG Lagrangians in which the lapse function is non-dynamical.
If we start with more general degenerate SCG Lagrangians (e.g., with a dynamical lapse function  \cite{Gao:2018znj,Gao:2019lpz,Lin:2020nro}), the method in this work may be used to search for more general ghostfree scalar-tensor theory with higher order derivatives and curvature terms. We shall report the progress in the future.

\acknowledgments

This work was partly supported by the National Natural Science Foundation of China (NSFC) under the grant No. 11975020.

%

\end{document}